\documentclass[prl,showpacs,floatfix,twocolumn]{revtex4}

\usepackage{graphicx}
\usepackage{amsmath}

\begin{document}

\title{Crossover of phase qubit dynamics in presence of negative-result weak measurement}

\author{Rusko Ruskov$^1$\footnote{On leave from INRNE, Sofia BG-1784, Bulgaria},
Ari Mizel$^1$, and Alexander N. Korotkov$^2$}

\affiliation{$^1$Department of Physics and Materials Research
Institute, Penn State University, University Park, Pennsylvania
16802, U.S.A. \\
$^2$Department of Electrical
Engineering, University of California, Riverside, California 92521,
U.S.A.}

\date{\today}

\begin{abstract}
   Coherent dynamics of a superconducting phase qubit is considered in the presence of
both unitary evolution due to microwave driving and continuous
non-unitary collapse due to negative-result measurement.
   In the case of a relatively weak driving, the qubit dynamics is dominated by the
non-unitary evolution, and the qubit state tends to an
asymptotically stable point on the Bloch sphere. This dynamics can
be clearly distinguished from conventional decoherence by tracking
the state purity and the measurement invariant (``murity''). When
the microwave driving strength exceeds certain critical value, the
dynamics changes to non-decaying oscillations: any initial state
returns exactly to itself periodically in spite of non-unitary
dynamics. The predictions can be verified using a modification of a
recent experiment.
\end{abstract}

 \maketitle

The problem of measurement of a single quantum system
plays a fundamental role in our understanding of physical reality \cite{Zurek}.
        While the evolution of an isolated quantum system is governed by its
Hamiltonian,
the state evolution of a measured (open) quantum system arises from
a non-trivial interplay between its internal Hamiltonian evolution
and the ``informational'' evolution associated with a given
measurement record \cite{Carmichael}.

        Experimental advances in the fabrication of
superconducting and semiconductor qubits \cite{superqubits} provide
unique possibilities to probe the quantum behavior of a single
quantum system by weakly measuring it via mesoscopic detectors.
For example, the qubit evolution can be monitored by a weakly coupled
quantum point contact or single-electron transistor, which behaves
classically on the time scale defined by the qubit dynamics.
        The measurement record in this case is a fluctuating current
\cite{Buks} that is correlated with the quantum state. Given the
continuous measurement record  the qubit state is continuously
collapsed due to quantum back action \cite{Kor-99-01,back-action}.

        Recently \cite{Science-Katz}, a variant of weak continuous measurement
was demonstrated experimentally in which partial collapse is
achieved by means of {\it registering  no signal}. Realized with a
superconducting ``phase'' qubit measured via tunneling
\cite{Cooper}, it is the first solid-state demonstration of quantum
null (negative-result) measurement effects proposed and discussed
mainly in the context of quantum optics \cite{Epstein-Dicke,
Dalibard}. Contrary to naive expectation, the {\it no signal} result
leads to a change of the quantum state, providing a new type of
qubit manipulation.

\begin{figure}
\vspace*{0.1cm}
\centering
\includegraphics[width=3.0in]{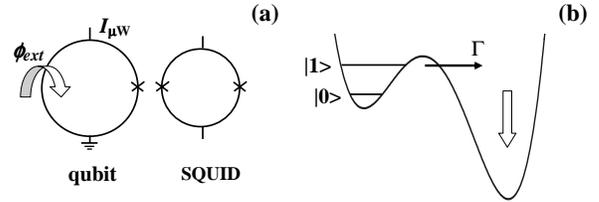}
\vspace{-0.06cm}
\caption{(a) Schematic of a phase qubit controlled by a microwave
current $I_{\mu W}$ and an external flux $\phi_{ext}$, and
inductively coupled to a SQUID. (b) Lowest energy levels in the left
well of the profile $V(\phi)$ represent the qubit states. Tunneling
to the right well from state $|1\rangle$ is detected by the SQUID.}
\label{schematic}
\end{figure}

        In this Letter we consider the interplay of coherent dynamics of
a phase qubit (see Fig.\ 1) due to unitary evolution (because of
microwave driving) and due to continuous collapse under a
negative-result measurement. We show existence
 of a {\it critical value} for the ratio of the Rabi frequency $\Omega_R$ to the
measurement rate $\Gamma$.
For $2\Omega_R/\Gamma <1$ the dynamics is dominated by the
 non-unitary evolution.  The qubit state is continuously collapsed to a fixed asymptotic state,
a special point on the Bloch sphere that depends on
$\Omega_R/\Gamma$ but is independent of the initial conditions.  Thus, every state purifies.  For
$2\Omega_R/\Gamma >1$, the dynamics changes qualitatively and shows
{\it non-decaying} oscillations of the qubit state,
so that no asymptotic state is reached and any initial
state returns to itself after an oscillation period. The qubit does not completely purify: the
purity undergoes non-decaying oscillations as well.

       We consider
a phase qubit \cite{Cooper,Science-Katz} which consists of a
superconducting loop interrupted by a Josephson junction [Fig.\
1(a)] and controlled by an external magnetic flux $\phi_{ext}$.
The qubit basis states $|0\rangle$ and $|1\rangle$ are the two
lowest energy states in the shallow ``left'' well [Fig.\ 1(b)] of
the potential profile $V(\phi)$, where $\phi$ is the superconducting
phase difference across the junction. A Rabi rotation of the qubit
state is achieved by applying a resonant microwave signal $I_{\mu
W}$. The measurement is performed by lowering the barrier (by
changing flux $\phi_{ext}$ that biases the qubit) for a finite time
$t$. While the tunneling from the ground state $|0\rangle$ is still
strongly suppressed, the excited state $|1\rangle$ may tunnel out,
with a rate $\Gamma$, to the much deeper right well where it
decoheres rapidly due to energy relaxation [Fig.\ 1(b)]. The
tunneling event can be detected by an inductively coupled SQUID
detector.

        First we consider the case when {\it no microwaves are applied.}
Measuring the qubit for sufficiently long time $t$, such that
$\Gamma t \gg 1$, essentially results in a
strong measurement: the qubit state
is either collapsed onto state $|0\rangle$ (if no tunneling has
happened) or destroyed (if tunneling has happened). However,
measurement for a finite time $t \lesssim \Gamma^{-1}$ is weak: the
qubit state is still destroyed if a tunneling event happens, but in
the case of no tunneling, a negative result, the qubit density
matrix $\rho$ evolves continuously according to quantum Bayesian
rule \cite{Kor-99-01,Science-Katz,Dalibard}:
\begin{eqnarray}
&& \rho_{00}( t) = \frac{\rho_{00}(0)}{ {\cal{N}} },\
\rho_{11}( t) = \frac{\rho_{11}(0)\, e^{-\Gamma t}}{ {\cal{N}} },   \label{Bayes-upd1}\\
&& \rho_{01}( t) = \rho_{01}(0)\,
\sqrt{\frac{\rho_{00}(t)\,\rho_{11}(t)}{\rho_{00}(0)\,\rho_{11}(0)}}
\ e^{ i \varphi} ,
\label{Bayes-upd2}
\end{eqnarray}
where ${\cal{N}} \equiv \rho_{00}(0) + \rho_{11}(0)\, e^{-\Gamma t}$.
Note that Eqs.\ (\ref{Bayes-upd1}) and (\ref{Bayes-upd2}) describe
the ideal change in $\rho$ in the rotating frame. The qubit acquires
a known phase $\varphi$ due to a small shift of the level spacing
under the change
of $\phi_{ext}$ \cite{Science-Katz}; in what follows we neglect this
effect assuming $\varphi=0$.
Possible decoherence effects will be discussed later.

        It proves convenient (Fig.\ 2) to characterize the quality of the qubit state by the
purity $P\equiv 2\mbox{Tr}\hat{\rho}^2 - 1 = x^2+y^2+z^2$ which is
an invariant of unitary transformations; here
$x=2\,\mbox{Re}\rho_{01}$, $y=2\,\mbox{Im}\rho_{01}$, and
$z=\rho_{00}-\rho_{11}$ are Bloch components. (Note that the linear
entropy, $1-P$, is a one-to-one function of the von Neumann entropy
$S=-\mbox{Tr}\hat{\rho}\log_2\hat{\rho}$, while $1-P\leq S$.)
Another important state characteristic is the ``murity''
$M=|\rho_{01}|^2/(\rho_{00}\,\rho_{11}) = (x^2+y^2)/(1-z^2)$
\cite{Kor-99-01,JordanKorotkov}, which is an invariant of the
measurement evolution [see Figs.\ 2(a,b)]. Obviously,
$P=M=1$ for a pure state, and it is easy to show that $P \geq M$
always.

\begin{figure}
\vspace*{0.1cm}
\centering
\includegraphics[width=3.4in]{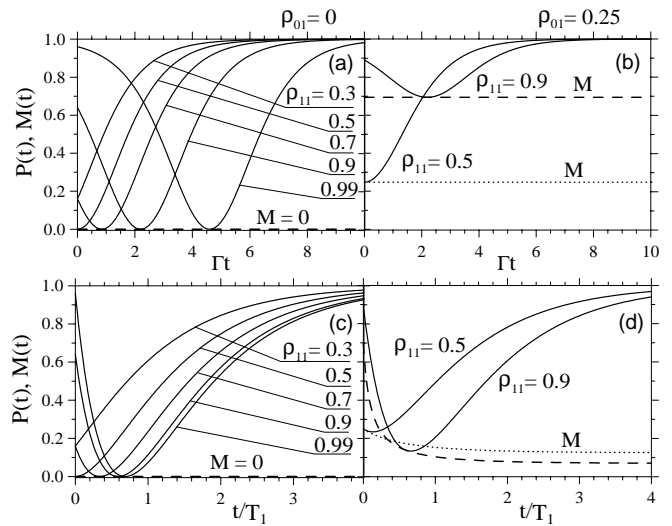}
\vspace{-0.15cm}
\caption{Time dependence of purity $P$ (solid lines) and murity $M$
(dashed or dotted lines) in the absence of microwave driving for
qubit state evolution due to (a,b)  negative-result measurement or
(c,d) zero-temperature energy relaxation. Notice qualitative
difference of the corresponding curves in the upper and lower
 panels for the same initial conditions [$\rho_{11}(0)$ is shown
for each curve, $\rho_{01}(0)=0$ in (a,c) and $\rho_{01}(0)=0.25$ in
(b,d)]. The relative scale of the upper and lower panels is chosen
to maximize visual similarity of curves in (a) and (c).
  } \label{PMevol-NullvsT1T2}
\end{figure}

From Eqs.\ (\ref{Bayes-upd1}) and (\ref{Bayes-upd2}) it is easy to
obtain the purity evolution due to negative-result measurement:
$P(t) = P(0) + \left[1-P(0)\right] ( 1 - e^{-\Gamma t}/ {\cal{N}}^2)$.
An initially pure state remains pure, while an initially mixed state
tends to the pure state $|0\rangle$  asymptotically, for
$t\gg 1/\Gamma$. However, the purity of an initially mixed state increases
monotonically in time only if $\rho_{11}(0) \leq 1/2$ [see Figs.\ 2(a,b)].
In contrast, for $\rho_{11}(0) > 1/2$ the state first
becomes more mixed, reaching
minimal purity $P_{min}^{meas} = M(0)$
at a time
    \begin{equation}
\tau_{min}^{meas} = \frac{1}{\Gamma}\,\ln \left(
\frac{\rho_{11}(0)}{1-\rho_{11}(0)} \right)
\label{t-min-null}
\end{equation}
when $\rho_{00}(t)=\rho_{11}(t)$,
 and only then starts to purify.

    A qualitative explanation of the nonmonotonic behavior of $P(t)$
is purely classical and based on the informational character of the
evolution due to negative-result measurement.
        Consider a (classical) qubit state which is known to be more likely
in state $|1\rangle$ than $|0\rangle$. If the qubit does not tunnel
out after a small time $t$, then the likelihood that the qubit is  
in the non-decaying state $|0\rangle$ slightly
increases. Thus, the uncertainty (entropy) of the qubit state
increases, and so the purity $P(t)$ decreases. If the qubit still
has not decayed after a sufficiently long time, we are practically
sure that the state is  
$|0\rangle$, so the entropy
decreases and purity increases. Notice that Eq.\ (\ref{t-min-null})
does not depend on $\rho_{01}(0)$, thus allowing a purely classical
interpretation.

    An important question is whether or not the qubit evolution
(\ref{Bayes-upd1})--(\ref{Bayes-upd2}) due to negative-result
measurement can be imitated by the evolution due to conventional
decoherence characterized by energy relaxation time $T_1$ and
dephasing time $T_2$ \cite{Weiss}. As we show below, the answer is
no, the two evolutions are significantly different.

Let us start the comparison assuming zero-temperature relaxation and
minimal dephasing rate, $T_2=2T_1$.
In the case $\rho_{01}(0) \ne 0$, the most obvious difference
between the two kinds of evolution is the behavior of the murity
$M(t)$ [see Figs.\ 2(b,d)].  The murity is a constant for the measurement
evolution, while in the case of decoherence $M(t)$ decreases to a
value $M(0) \rho_{00}(0)$, even though the qubit
approaches the same ground state.
If $\rho_{11}(0)>1/2$,   
then the minimal purity in the decoherence scenario,
$P_{min}^{T_1} = 1 - \left[1-|\rho_{01}(0)|^2/\rho_{11}(0)\right]^2$
is also smaller than $P_{min}^{meas} = M(0)$. This minimum is reached at a time
\begin{equation}
\tau_{min}^{T_1} = T_1\,\ln \{ 2\, \rho_{11}(0)/[\, 1 - |\rho_{01}(0)|^2/\rho_{11}(0)\, ] \}
\label{t-min-t1}
\end{equation}
that is much less sensitive than $\tau^{meas}_{min}$
[Eq.\ (\ref{t-min-null})]
to the initial conditions.
In particular, $\tau_{min}^{T_1}$ approaches the finite value
$T_1\ln 2$ for $\rho_{11}(0)\to 1$, while $\tau_{min}^{meas}$ grows
logarithmically. Notice [Figs.\ 2(b,d)] that in both evolutions the curve $P(t)$
touches the curve $M(t)$; in the measurement case this happens
at $\tau_{min}^{meas}$ while in the decoherence case this
happens at $t=T_1\ln 2\rho_{11}(0) < \tau_{min}^{T_1}$ when
$P(t)=M(t)=2M(0)\rho_{00}(0)$.  If $\rho_{01}(0)= 0$, Eqs. (\ref{t-min-null})
and (\ref{t-min-t1}) still imply a difference in their sensitivity to initial conditions
that is evident in Figs.\ 2(a,c).

When dephasing exceeds its minimal value ($T_2 < 2 T_1$), the
asymptotic murity $M(\infty )$ in the decoherence scenario always
tends to zero, thus making the two evolutions still more distinct.
Finite temperature leads to a similar effect and also makes the
asymptotic qubit state different from the asymptotic state
$|0\rangle$ of the evolution due to negative-result measurement.
Thus, the two evolutions are always significantly different.

Now let us consider the state dynamics due to negative-result
measurement {\it in the presence of microwave driving} exactly at
resonance. Differentiating Eqs.\
(\ref{Bayes-upd1})--(\ref{Bayes-upd2}) over time and adding the
evolution due to Rabi oscillations, we obtain the following
evolution in the rotating frame:
\begin{eqnarray}
\!\!\!\!\!&& \dot{\rho}_{00} = - \dot{\rho}_{11} = -\Omega_R\ \mbox{Im}\rho_{01}
+ \Gamma\, \rho_{00}\, \rho_{11}  \qquad\qquad\qquad\
\label{null-rho-evol1}
\\
\!\!\!\!\!&& \dot{\rho}_{01} =  i \frac{\Omega_R}{2}\
(\rho_{00} - \rho_{11}) - \frac{\Gamma}{2}\, (\rho_{00} - \rho_{11})\, \rho_{01}.
\label{null-rho-evol2}
\end{eqnarray}
Here we consider driving that shows up as a $\sigma_x$-term in the
Hamiltonian, ${\cal H}=\frac{\Omega_R}{2}(\left| 0 \right> \left< 1
\right|+\left| 1 \right> \left< 0 \right|)$; $\sigma_y$-evolution
can be easily incorporated via a finite rotation (for simplicity we
assume the absence of a $\sigma_z$-term). Though the equations
(\ref{null-rho-evol1}) and (\ref{null-rho-evol2}) are deterministic,
their non-linear terms resemble those in the case of noisy weak measurement
\cite{Kor-99-01}.

\begin{figure}
\vspace*{0.1cm}
\centering
\includegraphics[width=3.15in]{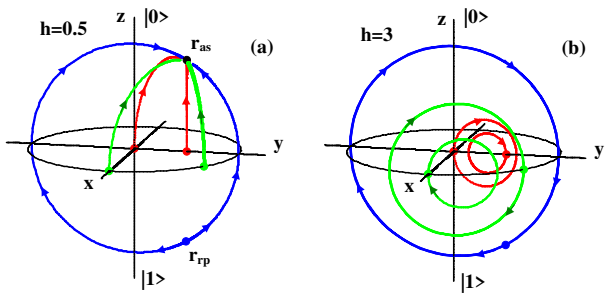}
\vspace{-0.25cm} \caption{
    (a) Trajectories of the qubit state evolution in Bloch coordinates
for $h\equiv 2\Omega_R/\Gamma =0.5$,
starting from several initial states. All states purify and approach
the attractive asymptotic point $r_{as}$, while $r_{rp}$ is the
repulsive point. (b) Evolution for $h=3$ from the same initial
states: non-decaying oscillations with period $4\pi/\Gamma
\sqrt{h^2-1}$.
    }
\label{crossover}
\end{figure}

The solution is conveniently expressed in terms of Bloch components.
The evolution of $x$ decouples, and setting $h\equiv
2\Omega_R/\Gamma$ we solve for $z$ and $y$. Introducing
$\omega\equiv (\Gamma/2)\sqrt{1-h^2}$ and $v_0 \equiv
\mbox{arctanh}\ \left[ \frac{\sqrt{1-h^2} z(0)}{1-h\, y(0)}
\right]$, we obtain
\begin{eqnarray}
x(t)=  \frac{x(0) \, [1-h^2] \cosh v_0 }{ [1-h \, y(0)] \cosh(\omega
t + v_0) - h [h-y(0)] \cosh v_0 } , \,\,
\label{x-meas}\\
\nonumber \\
 y(t) = \frac{ h [1-h\, y(0)] \cosh (\omega t + v_0) -
[h-y(0)] \cosh v_0  } { [1-h \, y(0)] \cosh(\omega t + v_0) - h
[h-y(0)] \cosh v_0 } , \,\,
\label{y-meas}\\
\nonumber\\
 z(t) =
\frac{\sqrt{1-h^2} \, [1-h\, y(0)]
\sinh (\omega t + v_0)}{ [1-h \, y(0)] \cosh(\omega t + v_0) - h
[h-y(0)] \cosh v_0 } . \,\, \label{z-meas}
\end{eqnarray}

  The most important observation is a critical point at $h = 1$ (independent
of $\omega$ and $v_0$). Below the critical value, $h < 1$, the
evolution is dominated by the measurement, and the qubit state
asymptotically collapses to a stable value on the Bloch sphere with
coordinates $x_{\infty}=0$, $y_{\infty} = h$, $z_{\infty} = \sqrt{1-h^2}$.
This occurs independently of the initial conditions.
The asymptotic state $r_{as} \equiv \{x_{\infty},y_{\infty}, z_{\infty} \}$
attracts the trajectories on the Bloch sphere
[see Fig.\ 3(a)], while the state
$r_{rp} \equiv \{x_{\infty},y_{\infty},-z_{\infty} \}$ repels trajectories.
   It is simple to visualize the dynamics starting from a point
on the great circle, $y^2 + z^2 = 1$. Then the presence of
microwaves rotates the state around the circle in a clockwise
direction (when viewed from the positive $x$ axis), while the
measurement evolution rotates it either clockwise or
counterclockwise towards state $|0\rangle$ (North Pole). At points
$r_{as}$ and $r_{rp}$ the two rotations exactly compensate each other,
creating the stable and unstable equilibrium states.

At the critical value, $h=1$, the equilibrium states $r_{as}$ and
$r_{rp}$ coincide, and the asymptote is achieved not exponentially, but
in a power-law fashion:
 $z(t)\simeq 4/(\Gamma t)$,
while $y(t)\simeq 1 - 8/(\Gamma t)^2$.
    Above the critical value, $h>1$, the state does not stabilize at
all, and the qubit performs {\it non-decaying oscillations} with
period $T_{osc}=4\pi/\Gamma \sqrt{h^2-1}$ [see Figs.\ 3(b) and 4].
This means that every state    
returns exactly to itself after $T_{osc}$, in spite of the
non-unitary dynamics \cite{objectivity} (thus being an example of
Quantum Un-Demolition measurement \cite{JordanKorotkov}).

\begin{figure}
\vspace*{0.1cm} \centering
\includegraphics[width=3.in,height=1.8in]{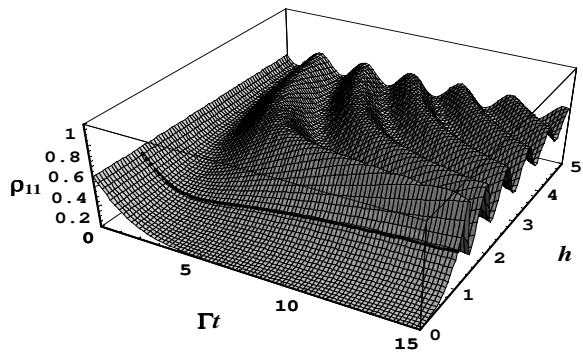}
\vspace{-0.1cm}
    \caption{Population of the excited state $\rho_{11}$
vs. scaled Rabi frequency $h$ and time $t$ for a totally mixed
initial state. Crossover from non-oscillatory to undamped oscillatory
dynamics occurs at $h=1$ (thick line). Decoherence is neglected.
    } \label{3D-plot}
\end{figure}

    The state purity evolves as $\dot{P} = \Gamma z\, (1-P)$, so a pure
state remains pure. For $h < 1$ a mixed state eventually purifies,
and asymptotically, $P \simeq 1 - e^{-z_{\infty}\Gamma t}$. Also, $M(t)$
eventually approaches $1$, though at a later time than the purity
($dM/dt \to 0$ for $h\to 0$). For $h>1$ the purity and murity
oscillate because of the state oscillation, so a mixed state does
not purify completely. In the case when $h$ is only a little over 1,
a mixed state purifies almost completely, but the purity still
returns to its initial value after a long period.

    Let us discuss now how, in the presence of microwaves,
negative-result measurement evolution differs from
decoherence evolution.
In the decoherence case, the measurement terms in
Eqs.\ (\ref{null-rho-evol1}) and (\ref{null-rho-evol2}) should be
replaced by    $-\gamma_1\, (\rho_{00} - p_{st})$ and $-\gamma_2\,
\rho_{01}$ respectively  \cite{Weiss}, where $\gamma_{1,2}\equiv
1/T_{1,2}$ and
$p_{st}$          
is the equilibrium ground state population (in experiment
\cite{Science-Katz,Cooper} $p_{st}\approx 1$.)
     Introducing $\tilde{h}\equiv 2\Omega_R/\gamma_1$ and $d \equiv
\gamma_2/\gamma_1 = T_1/T_2\geq 1/2$, we find
    \begin{eqnarray}
&& \hspace{-0.4cm} x(t)= x(0)\,\exp (-\gamma_2 t),
\qquad\qquad\qquad\qquad\qquad\qquad\qquad\ \,
\label{T1-evolx} \\
&& \hspace{-0.4cm} y(t) = y_{a} + \exp[-t(\gamma_1+\gamma_2)/2]
\nonumber\\
&& \hspace{0.2cm} \times \{ c_y \cosh{ \tilde{\omega} t } +
(\gamma_1/2\tilde{\omega}) [ \tilde{h} c_z + c_y (1-d) ]\, \sinh{
\tilde{\omega} t } \}, \qquad
\label{T1-evoly}\\
&& \hspace{-0.4cm} z(t) = z_{a} + \exp [-t(\gamma_1+\gamma_2)/2]
\nonumber\\
&& \hspace{0.2cm}\times \{ c_z \cosh{ \tilde{\omega} t } -
(\gamma_1/2\tilde{\omega}) [ \tilde{h} c_y + c_z (1-d) ]\, \sinh{
\tilde{\omega} t } \} , \label{T1-evolz}
\end{eqnarray}
where $\tilde{\omega}\equiv (\gamma_1/2) [(1 - d)^2 - \tilde{h}^2]^{1/2}$,
$c_z\equiv z(0) - z_a$, $c_y \equiv y(0) - y_a$, and
$\ x_a = 0$, $y_a = 2 (2p_{st}-1) \tilde{h}/(\tilde{h}^2 + 4d )$,
$z_a = 4 (2p_{st}-1) d /(\tilde{h}^2 + 4d )$ are the asymptotic values.

The evolution (\ref{T1-evolx})--(\ref{T1-evolz})
resembles that of a damped oscillator and is quite different from
the evolution (\ref{x-meas})--(\ref{z-meas}). For $\tilde{h}<|1-d|$
the overdamped regime is realized (no oscillations), while for
$\tilde{h}>|1-d|$ we have damped oscillations (underdamped regime).
For arbitrary $\tilde{h}$ the Bloch components approach $x_{a}$,
$y_{a}$, and $z_{a}$, and the asymptotic purity is $P_a =
4(2p_{st}-1)^2 (\tilde{h}^2 + 4d^2)/(\tilde{h}^2 + 4d)^2$ implying that a
mixed state never becomes pure except at zero temperature ($p_{st}=1$)
in the absence of microwaves. Even an initially pure state
becomes mixed, with asymptotic purity and murity both close to zero
for large $\tilde{h}$. We can conclude that in presence of
microwaves the qubit dynamics  due to decoherence is still
qualitatively different from the dynamics due to negative-result
measurement.

The phase qubit evolution due to negative-result measurement
discussed in this Letter can be verified experimentally using the
quantum state tomography \cite{Nori,Steffen} in the same way as in
the recent experiment of Ref. \onlinecite{Science-Katz}, which has verified the
evolution (\ref{Bayes-upd1})--(\ref{Bayes-upd2}).
    In a realistic situation the decoherence evolution is always
added to the measurement evolution; however, as we discussed above,
the qualitative effects of the two evolutions are easily
distinguishable. Moreover, the decoherence can be made more than 10
times slower than the evolution due to measurement
\cite{Science-Katz}, that justifies neglect of decoherence in our
analysis of the crossover from asymptotic qubit purification to
nondecaying oscillations.

    Notice that the predicted qubit evolution due to negative-result measurement
can be seen experimentally only as long as the qubit has not
decayed. The probability that the qubit has not decayed by time $t$
is ${\cal P}(t) = \exp [ -\Gamma\,\int_0^t dt' \rho_{11}(t') ]$.
In absence of microwaves ($h=0$) it becomes ${\cal
P}(t)=\rho_{00}(0)+\rho_{11}(0) e^{-\Gamma t}$ and remains finite
with increasing time. However, with added microwaves ($h\neq 0$)
${\cal P}(t)$ tends to zero at $t\to \infty$, which means that the
qubit eventually decays.
    Since predictions requiring unreasonably small
values of ${\cal P}$ are hardly accessible experimentally, we have
checked that the qualitative picture of our results can still be
seen at the cut-off level ${\cal P}>5\%$. While the close vicinity
of the critical point ($h=1$) is the hardest regime for experimental
analysis, the predicted non-oscillatory evolution at $h<1$ as well
as few non-decaying oscillations at $h\agt 3$ should be observable
experimentally with a minor modification of the experiment
\cite{Science-Katz}.

    The work was supported by the Packard foundation and DTO/ARO grant
W911NF-0401-0204.


\begin{thebibliography}{99}
\bibitem{objectivity} At $h > 1$ two observers with different
    knowledge of $\rho (0)$ will never agree with each other
    regarding the resulting state $\rho (t)$, in contrast to the
    case of non-destructive measurement \protect\cite{Kor-99-01}.
    However, the disagreement becomes less probable with time
    because of qubit decay.
\end{thebibliography}
\end{document}